\documentclass[graybox]{svmult}

\usepackage{mathptmx}       
\usepackage{helvet}         
\usepackage{courier}        
\usepackage{makeidx}         
\usepackage{graphicx}        
\usepackage{multicol}        
\usepackage[bottom]{footmisc}
\usepackage{natbib}

\begin{document}

\title*{A solution to the radiation pressure problem\\ in the formation of massive stars}
\author{Rolf Kuiper \and Hubert Klahr \and Henrik Beuther \and Thomas Henning}
\institute{
Rolf Kuiper 
\at 
Jet Propulsion Laboratory,
California Institute of Technology,
4800 Oak Grove Drive,
Pasadena, CA 91109,
USA, 
\email{Rolf.Kuiper@jpl.nasa.gov}
\and
Hubert Klahr, Henrik Beuther, Thomas Henning
\at
Max Planck Institute for Astronomy,
K$\ddot{\mbox{o}}$nigstuhl 17,
D-69117 Heidelberg,
Germany
}
\maketitle

\abstract{
We review our recent studies demonstrating that the radiation pressure problem in the formation of massive stars can be circumvented via an anisotropy of the thermal radiation field.
Such an anisotropy naturally establishes with the formation of a circumstellar disk.
The required angular momentum transport within the disk can be provided by developing gravitational torques.
Radiative Rayleigh-Taylor instabilities in the cavity regions -- as previously suggested in the literature  -- are not required and are shown to be not occurring in the context of massive star formation.
}

\vspace{-2mm}
\section{Introduction}
\label{sect:introduction}
\vspace{-2mm}
During their lifetime, massive stars exert a radiation pressure onto their surroundings, which is higher than their gravitational attraction.
How these massive stars can sustain accretion was an open question for decades.
Semi-analytical 
\citep{Kahn:1974p799}
and first radiation-hydrodynamical computations in spherical symmetry
\citep{Yorke:1977p376}
support the idea that the radiation pressure of a star retards the collapse of its proto-stellar core, leading to an upper stellar mass limit of 
$\le 40 \mbox{ M}_\odot$.

Based on the fact that in these 1D simulations the collapse was reversed by the re-emitted radiation, 
\cite{Nakano:1989p497}
inferred that the radiation pressure onto the accretion flow can be diminished by an anisotropic optical depth of the environment.
Such an anisotropy can be produced by the formation of an accretion disk.

The first numerical study aimed at verifying this idea was by 
\cite{Yorke:2002p735}.
In their simulations, the optical depth of the forming accretion disk turned out to be too low to provide a sufficient anisotropy of the radiation field.
As a result, the accretion onto the proto-star stopped shortly after the disk formation, limiting the maximum final stellar mass to $42.9 \mbox{ M}_\odot$.
In 
\cite{Krumholz:2009p687},
the authors claimed further feeding of the star and its circumstellar disk by a radiative Rayleigh-Taylor instability of the bipolar cavity regions.
The most massive (and still accreting) proto-star formed in this simulation was $41.5 \mbox{ M}_\odot$.

All in all, no star much more massive than the 1D radiation pressure barrier of $M_*^\mathrm{max} \approx 40 \mbox{ M}_\odot$ was formed in numerical simulations up to this point.

\vspace{-2mm}
\section{Methods}
\label{sect:methods}
\vspace{-2mm}
In this section, we highlight two of our code specifics, namely the grid in spherical coordinates and the hybrid radiation transport method developed.
For a complete description of the equations, methods, and numerical solvers please see 
\cite{Kuiper:2010p586}
for the radiation transport, 
\cite{Kuiper:2010p541}
for the hydrodynamics in axial symmetry, and 
\cite{Kuiper:2011p349}
for the hydrodynamics in three dimensions.

The grid of the computational domain is given in spherical coordinates with a logarithmically increasing resolution towards the center.
The resolution of the inner region around the centrally forming star is $\approx 1 \mbox{ AU}$, decreasing towards the outer core regions proportional to the radius.
This kind of grid covers the essential phenomena on the wide range of spatial scales from the outer core collapse (gravity dominated) over the torus formation (flattened rotating structure, still in-falling) in the transitional region down to the innermost disk formation (centrifugal balance).

The hybrid radiation transport method is highly adapted to the problem under investigation:
The stellar irradiation is computed via a very accurate frequency-dependent ray-tracing step, while the re-emission and thermal dust emission is computed within a fast gray Flux-Limited Diffusion (FLD) approach.

Hence, the feedback onto the directly irradiated regions around the massive proto-star -- the inner disk rim and the cavity walls -- is computed with very high accuracy, including the wide range of different optical depths for the broad stellar spectrum.
The ray-tracing of the stellar spectrum guarantees the resembling of the long range force of the irradiation, see 
\cite{Kuiper:2012p1151}.

\vspace{-2mm}
\section{Resolving the optical depth of the circumstellar accretion disk}
\vspace{-2mm}
In our simulation series for varying sizes 
of the inner sink cell 
\citep{Kuiper:2010p541},
we find that it is essential to fully include and resolve the innermost part of the dusty accretion disk, which depicts the highest optical depth, to compute the anisotropy of the thermal radiation field correctly.

Using the initial conditions of 
\cite{Yorke:2002p735},
in which the size of the sink cell is several times larger than the dust sublimation radius, our simulations resemble their results of short disk accretion epochs.
By restricting the sink cell to a size smaller than the dust sublimation radius, the thermal radiation field becomes strongly anisotropic and accretion sustains for several free-fall times.

\vspace{-2mm}
\section{Overcoming the radiation pressure barrier}
\vspace{-2mm}
In simulation series for varying initial core masses $M_\mathrm{core}$ in spherical as well as axial symmetry, we demonstrate, how the anisotropy of the thermal radiation field reduces the feedback onto the accretion flow, allowing the formation of the most massive stars known, see
\cite{Kuiper:2010p541}.
In spherical symmetry, the accretion is stopped and reversed by radiative forces.
Regardless of the initial core mass ($M_\mathrm{core} = 60 \ldots 480 \mbox{ M}_\odot$), the final mass of the forming star is limited to 
$< 40 \mbox{ M}_\odot$.
These results fully support the work by 
\cite{Kahn:1974p799}
and 
\cite{Yorke:1977p376}.

Setting the initial pre-stellar core in slow solid body rotation yields the formation of an accretion disk around the centrally forming massive star, which self-consistently leads to an anisotropy of the thermal radiation field.
The optical depth of the disk remains at high values by further accretion from the large scale envelope.
The bipolar cavity regions remain stable, 
and in turn lead to a mass loss of the proto-stellar core of roughly 50\% of the initial core mass.
In the shielded disk regions, the diminished radiative feedback onto the accretion flow enables the massive proto-star to increase its mass to $M_*^\mathrm{max} > 100 \mbox{ M}_\odot$.

In fact, these are the first simulations, including the effect of radiation pressure feedback, which demonstrate a possibility to form stars up to the maximum value of the observed stellar mass spectrum.

\vspace{-2mm}
\section{Angular momentum transport in massive accretion disks}
\vspace{-2mm}
In the 2D simulations we have to rely on an $\alpha$-viscosity model for the actual angular momentum transport.
In 
\cite{Kuiper:2011p349},
we demonstrate in a three-dimensional simulation 
that the self-gravity of the forming massive accretion disk self-consistently leads to gravitational instabilities and the formation of spiral arms, which gravitational torques in turn yield an angular momentum transport.
In contrast to the smooth viscous disk accretion in axial symmetry, the 3D simulation shows episodic accretion.
The accretion rate integrated over several episodic accretion events is as high as (even slightly higher than) the accretion rates of the axially symmetric simulations in 
\cite{Kuiper:2010p541}.
Hence, we conclude that the accretion rate required to form massive stars is self-consistently arranged by the self-gravity of the forming massive accretion disk.

\vspace{-2mm}
\section{On the radiative Rayleigh-Taylor instability}
\vspace{-2mm}
In 
\cite{Krumholz:2009p687}
the authors claim that a 3D radiative Rayleigh-Taylor instability in the bipolar cavity shell is required to allow further feeding of the star-disk system beyond the radiation pressure barrier.
Contrary, our simulations 
show the launching and expansion of stable outflow cavities, regardless of the dimension and for a large variety of initial conditions.
Among others, both studies differ in the treatment of the direct stellar irradiation feedback.
In 
\cite{Kuiper:2012p1151}
it is shown that if we cut down the radiation transport in our simulations to the gray FLD approximation, the forming cavities undergo a radiative Rayleigh-Taylor instability as well.
But if 
the stellar irradiation is computed with the much more sophisticated ray-tracing scheme, the cavities remain stable.
This result is backed up in 
\cite{Kuiper:2012p1151}
by analytical estimates of the radiative forces in the cavity shell, which are underestimated in the gray FLD approximation by up to two orders of magnitude.

\vspace{-2mm}
\section{Summary}
\label{sect:summary}
\vspace{-2mm}
In 45 published simulations regarding the radiation pressure feedback in the formation of massive stars, we have demonstrated a self-consistent and detailed description of the formation of the most massive stars known.
The radiation feedback in these simulations is computed via a sophisticated hybrid radiation transport approach, highly adopted to the problem under investigation, see 
\cite{Kuiper:2010p586}.
On the one hand, the simulation series fully recover the previous 1D radiation pressure barrier results of 
\cite{Kahn:1974p799}
and 
\cite{Yorke:1977p376}.
On the other hand, the simulation series attest the numerical and physical improvements in our studies of the radiation pressure feedback in the multi-dimensional context with respect to previous numerical studies in the field, see 
\cite{Kuiper:2010p541,Kuiper:2012p1151}.
The radiation pressure feedback onto the disk accretion flow is diminished by an anisotropy of the radiation field due to the forming massive accretion disk, 
see 
\cite{Kuiper:2010p541}.
The required angular momentum transport will self-consistently be provided by gravitational torques in the accretion disk, see 
\cite{Kuiper:2011p349}.

All in all, this demonstrates that the final masses of stars forming by accretion are not limited by the well-known radiation pressure barrier.

\begin{acknowledgement}
Author R.~K.~is currently financially supported by the German Academy of Science Leopoldina within the Leopoldina Fellowship programme, grant no.~LPDS 2011-5.
\end{acknowledgement}

\bibliographystyle{apalike}
\bibliography{Crete.bib}

\end{document}